\documentstyle[12pt,titlepage]{article} 
\title{QCD improved estimates to rare B-Decays
       at the heavy b-quark limit}
\author{Dongsheng Liu}          
\date{June, 1993}   
\def\_{\rule{.3em}{.15ex}}  

\begin{document}
\begin{titlepage}
 \begin{center}
  \vspace{0.75in}
  {\bf {\LARGE QCD improved exclusive rare $B$-Decays
               at the heavy b-quark limit } \\
  \vspace{0.2in}
   Dongsheng Liu$^\star$} \\
  International Centre For Theoretical Physics \\
  P.O. Box 586 \\
  34100 Trieste, Italy\\
  \vspace{0.5in}
  ABSTRACT \\
  \vspace{0.1in}
  \end{center}
  \begin{quotation}
  \noindent The renormalization effects from the b-quark scale down to the
            non-perturbative QCD regime are studied for rare $B$-decays at
            the heavy b-quark limit.
            Phenomenological consequences of these
            effects are investigated.
            We find that the anomalous scaling behavior plays a positive
            role in making non-perturbative model calculations consistent
            with recent CLEO measurements of $B\rightarrow K^* \gamma .$
\end{quotation}
\vspace{0.1in}
\begin{flushleft}
$\star$ {address after 1st Oct., 1993: Department of Physics,
        University of Tasmania, Hobart, Australia 7001}
\end{flushleft}

\end{titlepage}


\newcommand{\da}{\mbox{$\scriptscriptstyle \dag$}}
\newcommand{\lag}{\mbox{$\cal L$}}
\newcommand{\tr}{\mbox{\rm Tr\space}}
\newcommand{\fc}{\mbox{${\widetilde F}_\pi ^2$}}
\newcommand{\ns}{\textstyle}
\newcommand{\si}{\scriptstyle}

\section{Introduction}

Rare $B$-decays are vital testing grounds for the Standard Model and therefore
have
received a lot of theoretical attention \cite{RBD,GSW,AGM}.  Besides examining
the electroweak theory at the one-loop level, and providing quantitative
information on the yet undetermined top quark mass and CKM matrix elements
$V_{td}, V_{ts}, V_{tb}$, radiative rare $B$-decays could also be sensitive
to new physics beyond the Standard Model. Recently, the CLEO
group has reported the first observation of exclusive decay mode, $B\rightarrow
K^* \gamma$, and the measured average branching ratio, $BR( B \rightarrow
K^* \gamma)=(4.5 \pm 1.5 \pm 0.9) \times 10^{-5}$, as well as the inclusive
upper limit $BR( B \rightarrow X_s \gamma)< 5.4 \times 10^{-4}$, are consistent
with the Standard Model \cite{cleo}.

 To present theoretical estimates for exclusive processes, hadronic matrix
elements governed by nonperturbative physics have to be evaluated. A number of
effective approaches on the basis of symmetry considerations and
phenomenological
models for QCD in the nonperturbative regime have been developed and applied to
calculations of matrix elements for rare $B$-decays
\cite{ISGW,IW,ALT,DTP,DLT,OT,AOM,AL}.
As is well known, certain of
these methods, for example, the nonrelativistic quark model, cannot take into
account correctly the strong interaction effects at scales much larger than
the QCD scale $\Lambda_{QCD}$. For current operators containing the heavy
quark, the loop corrections with gluonic momenta between $\Lambda_{QCD}$
and $m_Q$ lead to large logarithms of the type $\alpha_s\ln \frac{m_Q}
{\Lambda_{QCD}}$. With renormalization group techniques, these large
logarithms can be summed to all orders, resulting in an anomalous scaling
factor. This kind of scaling factor has been worked out for the cases
such as heavy meson decay constants $f_M$, $B^0-\bar B^0$ mixing\cite{FB},
heavy meson semileptonic decay form factors\cite{FGGW}, and hadronic
$B$-decays \cite{DG}. In this
letter, we will consider the anomalous scaling behavior from the b-quark scale
down to the non-perturbative regime for current operators
appearing in rare $B$-decays upto next-to-leading order at the heavy
b-quark limit and present the consequences of this scaling for exclusive
rare $B$-decays.

                     In sect. II, we briefly review the renormalization
from $M_W$ to $m_b$ for rare decay operators in the full QCD with five quarks .
More or less, we follow a similar procedure in discussing
the anomalous scale below $m_b$, which is presented in sect. III.
The consequences of this QCD scaling for exclusive rare radiative
$B$-decays
are shown in sect. IV. We summarize in sect. V.

\section{ Renormalization in the full QCD}
To illustrate the general procedure of analyzing the anomalous scaling
behavior through the renormalization group techniques,
we briefly review the renormalization of current operators from heavy particle
scale, such as the top
quark and W boson, down to the bottom quark scale in QCD with five flavors.
At the moment, we ignore the running of the strong coupling constant between
$m_t$ and $M_W$. For the sake of concentration, we consider the rare radiative
bottom quark decay.
With no strong interaction in the form of QCD,  the effective hamiltonian for
the rare radiative bottom quark decay, after the t-quark and W-boson are
integrated out from the (1-loop) photonic pengiun diagrams, reads

 \begin{equation}
 {{\cal H} _{eff}} =- {{2G _F} \over {\sqrt 2}}(s_3 + s_2 e^{i\delta})
                  A(x_t)({{\alpha m_b} \over {4\pi}})
                   \bar s_L \sigma^{\mu \nu} b_R F_{\mu \nu},
\end{equation}
with $x_t=m_t^2/M_W^2$.
Here the contributions from virtual up and charm quarks have been ignored
and a term proportional to the strange quark mass
is dropped considering that $m_b \gg m_s$. $A(x)$ is the Inami-Lim function
with the form
\begin{equation}
A(x)=x \left [ \frac{ \frac{2}{3} x^2 + \frac{5}{12} x - \frac{7}{12} }
                    { (x-1)^3}
              -\frac{(\frac{3}{2} x^2 -x)\ln x}{(x-1)^4} \right ].
\end{equation}

The QCD effects play an important role in enhancing the radiative rare
$B$-decay, where the usual GIM suppresion factor in FCNC processes,
$\frac{m_q}{m_t}$, is replaced by a logarithmic one $\ln(\frac{m_t}{m_b})$.
Switching on strong interactions, the effective hamiltonian at $\mu \ll M_W$
is determined
by a proper operator basis. Following the notation of ref. \cite{GSW},
we have
 \begin{equation}
 {{\cal H} _{eff}} = {{4G _F} \over {\sqrt 2}} (s_3+s_2e^{i\delta})
                     {\displaystyle\sum^8_{j=1}} C_j(\mu) O_j(\mu) ,
\end{equation}
where
\begin{equation}\begin{array}{ll}\displaystyle
O_1=(\bar s_{L\alpha} \gamma^{\mu} b_{L\alpha})
    (\bar c_{L\beta} \gamma_{\mu} c_{L\beta}),    &
O_2=(\bar s_{L\alpha} \gamma^{\mu} b_{L\beta})
    (\bar c_{L\beta} \gamma_{\mu} c_{L\alpha}),   \\ \\

O_3=(\bar s_{L\alpha} \gamma^{\mu} b_{L\alpha})
    \sum_q (\bar q_{L\beta} \gamma_{\mu} q_{L\beta}),    &
O_4=(\bar s_{L\alpha} \gamma^{\mu} b_{L\beta})
    \sum_q (\bar q_{L\beta} \gamma_{\mu} q_{L\alpha}),   \\ \\

O_5=(\bar s_{L\alpha} \gamma^{\mu} b_{L\alpha})
    \sum_q (\bar q_{R\beta} \gamma_{\mu} q_{R\beta}),    &
O_6=(\bar s_{L\alpha} \gamma^{\mu} b_{L\beta})
    \sum_q (\bar q_{R\beta} \gamma_{\mu} q_{R\alpha}),   \\ \\

O_7=({{\alpha m_b} \over {4\pi}})
    \bar s_{L\alpha} \sigma^{\mu \nu} b_{R\alpha} F_{\mu \nu},  &
O_8=({{\alpha m_b} \over {4\pi}})
    \bar s_{L\alpha} \sigma^{\mu \nu} T^a_{\alpha\beta}
     b_{R\beta} G^a_{\mu \nu}.
\end{array}\end{equation}
with $\alpha$ and $\beta$ the color indices.

As the effective hamiltonian is renormalization scale independent, changes
of the coefficients with the scale should be compensated by changes of
these operators. This leads us to a renormalization group equation
\begin{equation}
{\displaystyle
\mu\frac{d}{d\mu}C_j(\mu)-\sum^8_{i=1} \gamma_{ij}(g)C_i(\mu)=0},
\end{equation}
where $\gamma$ is the anomalous dimension matrix of the operator basis
$\{O_j\}$.
Introducing the QCD $\beta$-function in terms of
\begin{equation}
{\displaystyle
\mu\frac{d}{d\mu}=\mu\frac{\partial}{\partial \mu}
                 +\beta(g)\frac{\partial}{\partial g}},
\end{equation}
the general solution of the above equation can be written in a matrix form as
\begin{equation}
{\displaystyle
{\bf C}(M_W/\mu, g(\mu))=\exp [\int^{g(\mu)}_{g(M_W)} dg
\frac{\gamma^T(g)}{\beta(g)}]
                 {\bf C}(1,g(M_W))}.
\end{equation}

Explicit solutions can be obtained perturbatively. In the leading
logarithmic approximation in which the terms like
$(\alpha_s\ln M_W/\mu)^n$ are summed to all orders, the $\beta$-function
and anomalous dimensions
calculated at the one-loop level of QCD are needed, while
matching conditions are of the zeroth order.  For the four-quark operators
$O_1 - O_6$, the matching conditions determined by the tree-level W-boson
exchange are
\begin{equation}
C_j(M_W)=0  \quad (j=1,3,4,5,6),
\end{equation}
\begin{equation}
C_2(M_W)=1.
\end{equation}
The values of $C_7(M_W)$ and $C_8(M_W)$ follow from (one-loop) penguin
diagrams.
They are
\begin{equation}
C_7(M_W)=-\frac{1}{2}A(x_t),
\end{equation}
and
\begin{equation}
C_8(M_W)=-\frac{1}{2}D(x_t),
\end{equation}
with
\begin{equation}
D(x)=\frac{x}{2} \left [ \frac{ \frac{1}{2} x^2 - \frac{5}{2} x - 1 }
                    { (x-1)^3}
              -\frac{3x\ln x}{(x-1)^4} \right ].
\end{equation}
Furthermore at the next-to-leading order, the $\beta$-function
and anomalous dimensions
are calculated to two loops of QCD and matching
conditions are of the first order. In this case, the logarithmic
terms like $\alpha_s (\alpha_s\ln M_W/\mu)^n$ are summed to all orders.

For the two-body
 decay $\bar B\rightarrow X_s \gamma$, which is modelled by $b\rightarrow s
\gamma$
 at the quark level,
 only the photon magnetic moment type operator $O_7$ contributes and the QCD
 corrected inclusive rare radiative decay rate is as follows
\begin{equation}
\Gamma (\bar B \rightarrow X_s \gamma)
= \frac{G_F^2 m_b^5\alpha}{32\pi^4}(s_2^2+s_3^2+2s_2s_3c_{\delta})
  \vert C_7(\mu) \vert^2
\end{equation}
Practically, this rate may be normalized to the semileptonic $B$-decay.
In this way we remove the quark mixing angles in the small mixing limit and
reduce the dependence on the bottom quark mass. So we have
\begin{equation}
\Gamma (\bar B \rightarrow X_s \gamma)
= \frac{6 \alpha}{\pi}\frac{\vert C_7(\mu) \vert^2}{f(m_c/m_b)}
\left ( 1 - \frac{2\alpha_s(m_b)}{3\pi} g(m_c/m_b) \right )^{-1}
\Gamma (\bar B \rightarrow X_c l \bar\nu_l),
\end{equation}
where $f(m_c/m_b)=0.45$ and $g(m_c/m_b)=2.4$ correspond to the phase space
factor and the one-loop QCD corrections to the semileptonic decay,
respectively.

In the leading logarithmic approximation and with an
anomalous dimension matrix for the truncated basis $O_1, \; O_2, \;
{\rm and}\; O_7$
\begin{equation}
\gamma =\frac{g^2}{16\pi^2} \left |
   \begin{array}{ccc}
   -1 &  3 & 0 \\  \\
    3 & -1 & 232/81 \\ \\
    0 &  0 & 16/3
    \end{array}
    \right |,
\end{equation}
 the renormalization coefficient
for the operator $O_7$ has the following analytic form\cite{GSW}
\begin{equation}
C_7(\mu)=- \eta^{16/23} \left \{ \frac{1}{2} A(x_t)
                              + [(58/135)(\eta^{-10/23}-1)
                                +(29/189)(\eta^{-28/23}-1)]
                        \right \},
\end{equation}
with $\eta = \alpha_s(M_W)/ \alpha_s(\mu)$.
Ref. \cite{GSW} has presented a discussion on the validity of the extra
approximation
with the truncated anomalous dimension matrix and estimated that the error in
this
coefficient function is less than 15\%. Furthermore the calculation of ref.
\cite{misiak} in the leading logarithmic approximation without the truncation
of the anomalous dimesion matrix also does not manifest significant change in
the above estimates.
Recently, the next-to-leading logarithmic effects which weaken the
QCD-corrections
to the $b\rightarrow s \gamma$, have been included partly in ref.
\cite{misiak2} and result in a change of 10-20\% in the inclusive decay rate.

We notice that the upper bound of theoretical estimate for the inclusive
rate is relevant to the renormalization scale which is usually
chosen about the b-quark mass.  Actually, based on eq. (14) and (16),
a renormalization scale extrapolated almost to $2.0GeV$ is used to set the
upper
bound of the estimate\cite{AGM}

\begin{equation}
BR(\bar B \rightarrow X_s + \gamma) = ( 2 \sim 5 )\times 10^{-4}.
\end{equation}
in the minimal standard model.

\section{Anomalous scaling at the heavy b-quark limit}

 In this section, we present an analysis for
renormalization of operators considered in the previous section
below the bottom quark mass. There are three flavors of quarks in the area
concerned.
The masses of bottom and strange quarks lie somehow on the
upper and lower bounds,
respectively, whilst the charm quark locates in the middle. Based on this
observation, we will scale down by treating the bottom quark
as a heavy particle and integrating it in using an effective theory with
four flavors.
However, when we proceed into the regime between $m_c$ and $m_s$,
where both bottom and charm quarks are heavy, the light flavor number becomes
three.

Working in an effective theory
where the bottom quark is treated as a heavy particle, we have the following
expansion for $O_j$
\begin{equation}
O_j\cong\displaystyle\sum_l D_{jl}(m_b/\mu, g(\mu)) O_l^{\prime}(\mu)
 + {\cal O}(\frac{1}{m_b}),
\end{equation}
where $\{O_l^{\prime}\}$ is an operator basis of the lowest possible dimension.
Contributions of higher dimension operators vanish at the
heavy bottom quark limit. Operators $O_l^{\prime}$ only depend upon the light
scale $\mu$ and the large logarithm $\ln \frac{m_b}{\mu}$ is transfered into
the cofficients $D_{jl}$, which obey the following renormalization group
equations
\begin{equation}
{\displaystyle
[ \mu\frac{\partial}{\partial\mu}
 +\beta(g)\frac{\partial}{\partial g}] D_{jl}(\mu)
-\left [ \sum_k \gamma^{\prime}_{kl}(g^{\prime})D_{jk}(\mu)
       - \sum_i \gamma_{ji}(g) D_{il}(\mu) \right ] = 0},
\end{equation}
where apart from different light flavor counting, the QCD $\beta$-function
retains the same form as the scale goes down, namely
\begin{equation}
\beta(g) = -g\left [\beta_0  \frac{g^2}{16\pi^2}+\beta_1
(\frac{g^2}{16\pi^2})^2
             + \cdots \right ] ,
\end{equation}
with the one-loop (two-loop) coefficient listed in Table 1. Along with
$\gamma^{\prime}$, the anomalous dimensions of $\{O_l^{\prime}\}$ in the
heavy quark effective field theory(HqEFT),
there is also a term in the above equation associated with $\gamma$, the
amomalous dimensions of $O_j$
in the higher scale theory. As shown explicitly in the appendix, the
combination
of the coefficients $C_j$ in eq. (7) with the part of $D_{jl}$
associated to the anomalous dimensions $\gamma_{ji}$ gives one the anomalous
scaling from $M_W$ down to $m_b$. We will drop the $\gamma$-term in eq. (19)
when we work below the bottom quark mass.

Since the photon magnectic moment
type operator $O_7$ dominates the rare radiative decay $b\rightarrow s \;
\gamma$, we concentrate on it at the present stage.
Let us consider the proper operators of dimension three \footnote[1]{ At the
moment we omit the phton field in $O_7$ and the bottom quark
mass that serves as the matching scale and is going to be set by the solution
of $\bar m(m_b)=m_b$, with $\bar m(\mu)$ the running mass.} in the effective
theory where, to remove the heavy mass, the bottom quark field is redefined
by
\begin{equation}
h_v(x)=e^{im_b v\cdot x \not v}\psi_b(x).
\end{equation}
Here $v_{\mu}$ is the four-velocity of the heavy bottom quark, which is a
conserved quantity with respect to the low-energy QCD.
Obviously, the first operator should be $O^{\prime}_1=\bar s_L \sigma_{\mu\nu}
h_{v R}$.
Constructing from the Dirac matrix $\gamma_{\mu}$ and the bottom quark velocity
which is an external parameter also
gives us the antisymmetric current operator  $O^{\prime}_2=\bar s_L
(\gamma_{\mu}v_{\nu}-v_{\mu}\gamma_{\nu})h_{v R}$.
Because the large
logarithmic term arises only from the vertex of the heavy quark with the gluon,
which becomes $ig\lambda^a v^{\mu}$ in the heavy quark limit,
the renormalization is independent
of the spin structure of the inserted current operators and does not mix them
up. In this case, the
anomalous dimension matrix reduces to a constant one and the renormalization
group equation for the coefficients has a quite simple form
\begin{equation}
{\displaystyle
\left [ \mu\frac{\partial}{\partial\mu}
       +\beta(g^{\prime})\frac{\partial}{\partial g^{\prime}}
       - \gamma^{\prime}(g^{\prime})
\right ] D_{7l}(\mu)} = 0.
\end{equation}
Expressing the solution of this equation in the same manner as eq. (7) gives
\begin{equation}\displaystyle
D_{7l}
=\exp \left [ \int^{g^{\prime}(m_c)}_{g^{\prime}(m_b)} dg^{\prime}
             \frac{\gamma^{\prime}(g^{\prime})}{\beta^{\prime}(g^{\prime})}
             +\int^{g^{\prime\prime}(\mu)}_{g^{\prime\prime}(m_c)}
                   dg^{\prime\prime}
             \frac{\gamma^{\prime\prime}(g^{\prime\prime})}
             {\beta^{\prime\prime}(g^{\prime\prime})}
             \right ] D_{7l}(m_b).
\end{equation}
The second part of this solution arises when we cross the charm quark mass
below which the light flavors reduce to three.
The charm quark is only involved as a virtual heavy particle in loops and
except different QCD couplings and light
flavor numbers being used,
$\gamma^{\prime\prime}$ is the same as $\gamma^{\prime}$ which has the form
\begin{equation}
\gamma^{\prime} = \gamma_0^{\prime}  \frac{g^{\prime 2}}{16\pi^2}
        +\gamma_1^{\prime} (\frac{g^{\prime 2}}{16\pi^2})^2
        + \cdots ,
\end{equation}
with the so-called ``hybrid ''anomalous dimension of ref. \cite{FB}
and the two-loop coefficient \cite{JING} shown in Table 1.

As a first step we work in the leading logarithmic approximation and get the
solutions
\begin{equation}
D_{71}(\mu)= \left [ \frac{\alpha_s^{\prime}(m_b)}{\alpha_s^{\prime}(m_c)}
              \right ]^{-\frac{6}{25}}  \quad
              \left [ \frac{\alpha_s^{\prime\prime}(m_c)}
                           {\alpha_s^{\prime\prime}(\mu)}
              \right ]^{-\frac{2}{9}} ,
\end{equation}
and
\begin{equation}
D_{7l}(\mu)=0, \quad \quad {\rm for}\; l\not= 1.
\end{equation}
Here the matching conditions $D_{71}(m_b)=1$, and $D_{7l}(m_b)=0$ for
$l\not= 1$ have been used. As matrix elements are concerned, the factor
containing
$\mu$ should be cancelled by the $\mu$-dependence of hadronic states which
are relevant to nonperturbative effects. This leaves us the anomalous heavy
quark mass
dependence
\begin{equation}
\Omega^T_{LLA}= \left [ \frac{\alpha_s^{\prime}(m_b)}{\alpha_s^{\prime}(m_c)}
              \right ]^{-\frac{6}{25}}  \quad
              \left [ \alpha_s^{\prime\prime}(m_c)
              \right ]^{-\frac{2}{9}}.
\end{equation}

Then we consider next-to-leading logarithmic contributions. Using the matching
condition for the current $\bar s \Gamma b$ to the $g^2$ order\cite{FB,FGGW}
\begin{equation}\displaystyle
\Gamma \longrightarrow [ \Gamma - \frac{g^2(m_b)}{24\pi^2}
                                  \gamma^{\lambda} \not v \Gamma \not v
                                  \gamma^{\lambda} ],
\end{equation}
we find that at $\mu=m_b$
\begin{equation}
\sigma_{\mu\nu}(\gamma_5)\longrightarrow\sigma_{\mu\nu}(\gamma_5),
\end{equation}
just as at the zeroth order up to a vertex renormalization factor. Thus
we also have only one non-zero coefficient, {\sl i.e.} $D_{71}$ at the
next-to-leading order. Once again we expand this coefficient function at $m_b$
in the QCD
coupling
\begin{equation}
D_{71}(m_b) = 1+d_1 \frac{g^{\prime 2}}{16\pi^2} + \cdots ,
\end{equation}
and find the next-to-leading correction to the heavy mass dependence
\begin{equation}\displaystyle
(d_1+\kappa^{\prime})\frac{\alpha_s^{\prime}(m_b)}{4\pi}
-\kappa^{\prime}\frac{\alpha_s^{\prime}(m_c)}{4\pi}
+\kappa^{\prime\prime}\frac{\alpha_s^{\prime\prime}(m_c)}{4\pi}.
\end{equation}
Here a $\mu$-dependence factor that is absorbed into the matrix element is
understood. It is worthwhile to point out that the combination
$d_1+\kappa^{\prime}$ is renormalization-scheme independent.
The $\kappa$-parameters determined by
\begin{equation}\displaystyle
\kappa=\frac{\gamma_0}{2\beta_0}
      (\frac{\gamma_1}{\gamma_0} - \frac{\beta_1}{\beta_0}),
\end{equation}
are listed in Table 2.
Matching the vertex renormalization to the one loop in the full theory
\cite{Neubert0} with that in the
effective theory \cite{FGGW,MND}
\begin{equation}
(Z_{full}-1)=(Z_{eff}-1)+\frac{g^{\prime 2}}{16\pi^2}d_1
\end{equation}
at $\mu=m_b$ in the ${\overline {MS}}$ scheme gives us $ d_1=-6c_F$, with
$c_F=4/3$.

Finally let us combine eq.(31) with eq. (27) and present the anomalous heavy
quark mass dependence for the photon magnetic moment type operator $O_7$ with
both of the leading and the next-to-leading logarithmic contributions
\begin{equation}\displaystyle
\Omega^T=
 \left [ \frac{\alpha_s^{\prime}(m_b)}{\alpha_s^{\prime}(m_c)}
          \right ]^{-\frac{6}{25}}  \;
          \left [ \alpha_s^{\prime\prime}(m_c)
          \right ]^{-\frac{2}{9}}
          \left [1 - 0.710 \alpha_s^{\prime}(m_b) + 0.073\alpha_s^{\prime}(m_c)
                   - 0.060\alpha_s^{\prime\prime}(m_c)
          \right ].
\end{equation}
The QCD fine structure constant accurate to the second order is
\begin{equation}\displaystyle
\alpha_s(m^2)=\frac{4\pi}{\beta_0 \ln(m^2/\Lambda^2_{\overline {MS}})}
              \left [1 -\frac{\beta_1 \ln\ln(m^2/\Lambda^2_{\overline {MS}})}
                             {\beta_0^2 \ln(m^2/\Lambda^2_{\overline {MS}})}
               \right ].
\end{equation}
As the heavy quark mass is concerned, we use the running mass $\bar m(\mu)$ in
the $\overline {MS}$ scheme and take $m_Q$ as the solution of $\bar m(m_Q)=
m_Q$. In terms of the invariant mass $\hat m$ we have the following form
\cite{Narison}
\begin{equation}\displaystyle
\bar m(\mu)= \hat m
             \left [ \ln (\mu/ \Lambda_{\overline {MS}}\right ]
                  ^{-\frac{\gamma_{m0}}{2\beta_0}}
            \left [ 1 - \frac{\gamma_{m0}}{2\beta_0}
                        \frac{\beta_1 \ln\ln(\mu^2/\Lambda^2_{\overline {MS}})}
                             {\beta_0^2 \ln(\mu^2/\Lambda^2_{\overline {MS}})}
                       + \frac{\kappa_m}
                              {\beta_0 \ln(\mu^2/\Lambda^2_{\overline {MS}})}
            \right ],
\end{equation}
where $\gamma_{m0}$ ($\gamma_{m1}$) is the one-loop (two-loop) coefficient of
the mass anomalous dimension (see Table 1). The values of $\kappa_m$ in terms
of eq. (32) are given
in Table 2. With invariant masses in ref. \cite{Narison} and $\Lambda_
{\overline {MS}}$ equal to $0.25\; GeV$, we obtain scale masses in the
${\overline
{MS}}$ scheme, $m_b=4.39\; GeV$ and $ m_c=1.32\; GeV.$  Fine structure
constants in $\Omega^T$ are
\begin{equation}
\alpha_s^{\prime}(m_b)= 0.204,\quad \alpha_s^{\prime}(m_c)= 0.332,
\quad \alpha_s^{\prime\prime}(m_c)=0.300,
\end{equation}
which give the following value
\begin{equation}
\Omega^T \cong 1.12\times 1.31\times (1-0.139) \cong 1.26
\end{equation}

Several comments are in order:

$\star \quad$ The heavy quark mass dependence in the leading logarithmic
approximation
is universal to the current operators $\bar s \Gamma b$ with $\Gamma$ any
matrix
in the Dirac space. However, the $\Gamma$-dependence in the full theory enters
into the effective theory at higher orders through matching conditions.

$\star$ Next-to-leading corrections are twofold. One comes from two-loop
anomalous dimensions and $\beta$-function and does not depend on $\Gamma$.
Another arises from matching conditions at the one-loop level and impacts on
the correction differently for different $\Gamma$. For instance, the
counterpart of $d_1$ in eq. (30) for vector and axial current operators
is $-3c_F$ \cite{JING,MND}. The large value of $d_1$ for the present case
leads to the result that the next-to-leading corrections
are dominated by the $\alpha_s^{\prime}(m_b)$ term in eq. (34) that weakenes
the leading effects.

$\star$ As we know, the photon magnetic moment type operator does not mix
into four quark operators and the gluon magnetic moment type operator. The QCD
corrections for the case in hand still remain unchanged when the mixings
of these operators with $O_7$ are taken into account.  But these mixings
add extra contributions, which have not been touched in this work and are
believed not to change the essential feature of the anomalous mass factor.
\section{ Phenomenological consequence}

When a non-perturbation method at the QCD scale is used to evaluate hadronic
matrix elements, the heavy-quark-mass dependence arising from renormalization
should be taken into account. A number of such methods,
such as the constituent
quark model (CQM), MIT Bag model, and the heavy quark limit (HQL),
have been employed in calculating rare $B$-decay matrix elements
\cite{ALT,DLT,OT,AOM,AL}. As a phenomenological application of our result
in the previous section, we modify these calculation by incorporating the
anomalous scale below the bottom quark mass. Our improved estimates
for the exclusive rare radiative $B$-decays are listed in Table 3.
The $R_0$ and $R^{\prime}$ are the ratio of exclusive to inclusive decay
widths, {\sl i.e.}
\begin{equation}\displaystyle
R=\frac{\Gamma(\bar B\rightarrow K^* \gamma)}{\Gamma(\bar B\rightarrow X_s
\gamma)},
\end{equation}
before and after the modification, respectively. The top
quark mass dependence and the coefficient $C_7$ are removed in this ratio.
In the pure CQM and MIT Bag
model calculations, the values of $R_0$ are about 0.05. This yields an
exclusive branching ratio of $Br(\bar B\rightarrow K^* \gamma)=2.1\times
10^{-5}$, for a top quark mass of $160$ GeV. It is small comparied to
the mean value of recent CLEO measurements. However, the anomalous
scale factor of LLA increases this ratio almost upto 0.11, but the
next-to-leading
correction weakens this enhancement and results in a value of 0.082 for the
improved ratio, which gives a branching ratio of $3.9\times 10^{-5}$. We find
that incorporating the anomalous scale mass factor
plays a positive role in making CQM and bag model calcualtions consistent
with the CLEO date. Using the range of $3.5\sim 12.2$ for $R$ in
ref.\cite{AOM} obtained through varing the oscillator strength
$\beta$ between $\beta_K=0.34\; GeV$ and $\beta_B=0.41\; GeV$,
we get a range of $(1.7\sim 5.7)\times 10^{-5}$ for the branching ratio of
$\bar B\rightarrow K^* \gamma$ decays.  Our numerical results are slightly
different
from that of ref.\cite{AOM} because of a different top quark mass being used.
This range is lifted
upto $(2.7\sim 9.4)\times 10^{-5}$ by the anomalous mass factor. The improved
estimates are completely in agreement with preliminay CLEO obseravtions.

Even though the strange quark is not particular heavy compared to the QCD sale,
HQL also presents reasonable estimate for the rare radiative $B$-decay.
We hope improvements will be included along this approach in the future.
As the authors of ref. \cite{OT} point out,  the large $R$ at the third row
of Table 3 is due to the use of a nonrelativistic recoil momentum. From
the theoretical point of view, this is quenstionable considering that
the recoil in the $B$ to $K^*$ decay is very large. Experimentally, estimates
with this large $R$ for rare radiative $B$-decays are not favored.

\section{Summary}

In this letter, we have made an analysis of the anomalous heavy quark mass
dependence of the rare $B$-decays by considering the anomalous scale of
current operators below the bottom quark mass. Combining this
anomalous mass factor evaluated upto next-to-leading order with
calculations using non-perturbative models, such as CQM
and MIT Bag Model, gives us exclusive rare radiative $B$-decay
rates which are in excellent agreement with the recent CLEO measurements.
On the other hand we observe that HQL also works as a
preliminary approximation to $B\rightarrow K$ processes.

When we work below the bottom quark mass in this letter, mixings of four
quark operators with the photon magnectic moment type operator have not been
touched. We expect that they do not change the essential feature of the
anomalous mass factor. The investigation of these mixings and their
phenomenological consequences is under way and will be presented elsewhere.

\noindent
{\bf Acknowledgement}

\noindent
The author thanks Professor A. Ali for many encouraging conversations.
He would like to thank his colleagues, Professor F. Hussain, Dr. G. Thompson,
R. Ahmady and Zhijian Tao for discussions. He also would like to thank
Dr. M. Neubert for many helpful communications.

\newpage

\noindent{\large {\bf Appendix A}}

In this appendix, we present the formal solution of the renormalization group
equations, eq. (5) for $\mu < M$ and eq. (19) for $\mu < m$ in the main text.
We will show that the renormalization below $m$ is effectively irrelevant to
the anomalous dimension matrix of operators $O_j$.

Once we  write down the effective hamiltonian in
terms of an operator basis
$$\displaystyle
{\cal H}_{eff} = \sum_j C_j O_j
\eqno(A.1)$$
at $\mu\ll M$, the coefficient ought to satisfy the renormalization group
equation, in a matrix form,
$$
{\displaystyle (\mu\frac{d}{d\mu}- \gamma^T){\bf C} =0},
\eqno(A.2)$$
where $\gamma$ is the anomalous dimension matrix. Introducing a transformation
matrix $V$ in such a way that
$$
(V^{-1}\gamma^T V)_{ij}=\tilde\gamma_j \delta_{ij},
\eqno(A.3)$$
where $\tilde\gamma_j$ are eigenvalues of the transpose of $\gamma$, we may
construct an alternate operator basis
though the linear combination $ \tilde {\bf O} =V^T {\bf O}$.
The effective hamiltonian can be rewritten in the same structure
$$
\displaystyle {\cal H}_{eff} = \sum_j \tilde C_j \tilde O_j.
\eqno(A.4)$$
The advantage of this basis is that operators are multiplicatively renormalized
and coefficients have solutions like
$${\displaystyle
\tilde C_j = \exp [\int^{g(\mu)}_{g(M)} dg \frac{\tilde\gamma_j}{\beta}]
                 \tilde C_j(M)}.
\eqno(A.5)$$

In the region $\mu\le m \le M$ where the proper operator basis is
\{ $O^{\prime}_l$ \}, there is the operator expansion
$$\displaystyle
O_j=\sum_l D_{jl} O^{\prime}_l,
\eqno(A.6)$$
and the corresponding renormalization group equation has the form
$${\displaystyle
(\mu\frac{d}{d\mu}+\gamma ){\bf D}_l-\sum_k \gamma^{\prime}_{kl}{\bf D}_k = 0},
\eqno(A.7)$$
with $\gamma^{\prime}$ the anomalous dimension matrix of $O^{\prime}_l$.
 Using the transform (A.3) and similarly
$$
(W^{-1}\gamma^{\prime T} W)_{kl}=\tilde\gamma_l^{\prime} \delta_{kl},
\eqno(A.8)$$
with $\tilde\gamma_l^{\prime}$ the eigenvalues of the transpose of
$\gamma^{\prime}$, the coefficients have the following solutions
$$\displaystyle
\tilde D_{jl}
=\exp \left [ -\int^{g(\mu)}_{g(m)} dg\frac{\tilde \gamma_j}{\beta}
              +\int^{g^{\prime}(\mu)}_{g^{\prime}(m)} dg^{\prime}
                     \frac{\tilde \gamma^{\prime}_l}{\beta^{\prime}}
      \right ] \tilde D_{jl}(m).
\eqno(A.9)$$
In terms of these coefficients, the effective hamiltonian becomes
$$\displaystyle
{\cal H}_{eff}=\sum_{jl} \tilde C_j \tilde D_{jl} \tilde O^{\prime}_l
              =\sum_{jl} \left \{ \exp \left [
               \int^{g(m)}_{g(M)} dg\frac{\tilde \gamma_j}{\beta}
                \right ] \tilde C_j (M)
                \exp \left [
                \int^{g^{\prime}(\mu)}_{g^{\prime}(m)} dg^{\prime}
                     \frac{\tilde \gamma^{\prime}_l}{\beta^{\prime}}
                             \right ] \tilde D_{jl}(m) \right \}
                             \tilde O^{\prime}_l.
\eqno(A.10)$$
It is remarkable that the combination of the first piece of $\tilde D_{jl}$,
which is associated with the anomalous dimensions in the effective theory
above $m$, with $\tilde C_j$ in eq (A.5) cancels the $\mu$-dependence and
gives the anomalous scaling behavior from $M$ down to $m$, which is
represented by the first coefficient of the effective hamiltonian in eq.
(A.10).
The second coefficient, representing the anomalous scale from $m$ down to
$\mu$, is determined by the anomalous dimensions in the effective theory at
$\mu\le m$.
\newpage
Table 1. Perturbation coefficients for the $\beta$-function
and anomalous dimensions

\begin{tabular}{c|c|c|c|c|c} \hline
\multicolumn{2}{c|}{$\beta$-function}        &
\multicolumn{2}{c|}{anomalous dimensions}    &
\multicolumn{2}{c}{anomalous dimensions}    \\
\multicolumn{2}{c|}{}                                     &
\multicolumn{2}{c|}{of the quark mass }                   &
\multicolumn{2}{c}{of $\bar s \Gamma b$ in the HqEFT}     \\
\hline
$\beta_0$              & $\beta_1 $           &
$\gamma_{m0}$          & $\gamma_{m1}$        &
$\gamma^{\prime}_0$    & $\gamma^{\prime}_1$  \\
\hline
& & & & & \\
$11-\frac{2}{3}N_f$    & $102-\frac{38}{3}N_f$             &
      8                & $\frac{8}{3}(\frac{101}{2}
                                     -\frac{5}{3}N_f)$      &
     -\,4                & $-(\frac{254}{9}+\frac{56}{27} \pi^2
                           -\frac{20}{9}N_f)$              \\
& & & & & \\  \hline
\end{tabular}

\vskip 5pc
Table 2. Summary of values for the $\kappa$-parameter

\begin{tabular}{c|c|c|c} \hline
\multicolumn{2}{c|}{$N_f=4$}  & \multicolumn{2}{c}{$N_f=3$} \\  \hline
$\kappa^{\prime}$ & $\kappa^{\prime}_m$ & $\kappa^{\prime\prime}$ &
$\kappa^{\prime\prime}_m$ \\  \hline  & & & \\
$\frac{199}{625}-\frac{28}{225}\pi^2 (\cong -0.909)$ &
$\frac{7606}{1875} (\cong 4.06)$ &
$\frac{31}{81}-\frac{28}{243}\pi^2 (\cong -0.753)$ &
$\frac{290}{81} (\cong 3.58)$ \\
& & & \\  \hline
\end{tabular}

\vskip 5pc
Table 3. Improved branching ratios for $\bar B \rightarrow K^* \gamma$

\begin{tabular}{c|c|c|c|c} \hline
Model   & ref.       & $R_0$(\%) & $R^{\prime}$(\%) &
$Br(B\rightarrow K^* \gamma)\times 10^5$     \\  \hline
CQM(r)  & \cite{ALT}  & $4.5$     & $7.4$  & $3.5$              \\ \hline
CQM \& MIT Bag & \cite{DLT} & $6.0$ & $9.8$ &  $4.6$       \\ \hline
CMQ(nr) & \cite{OT}  & $21.$     & $ 34.$ &       $16.1$        \\ \hline
HQL+CQM & \cite{AOM} & $3.5\sim 12.2$ & $5.7\sim 20.0$ & $2.7\sim 9.4$ \\
\hline
HQL     & \cite{AL} & $12.\sim 27.$  & $20.\sim 44.$  & $9.4\sim 20.5$ \\
\hline
\end{tabular}

\end{document}